\documentclass[aip,graphicx]{revtex4-1}

\usepackage{graphicx}
\usepackage{dcolumn}
\usepackage{bm}
\usepackage{mathtools}

\draft

\begin{document}

\title
{Interaction between a Waveguide Mode and an Absorptive Medium}

\author{Hadi Amarloo}
\email{hamarloo@uwaterloo.ca}
\author{Safieddin Safavi-Naeini}
\affiliation{Department of Electrical and Computer Engineering, University of Waterloo, \\ Waterloo, ON, Canada}

\begin{abstract}
\vspace{0.75cm}
\textbf{Abstract}-- Absorption spectroscopy studies the absorption of electromagnetic wave in a material sample under test. The interacting electromagnetic wave can be propagating in free space or in a waveguide. The waveguide-based absorption spectroscopy method has several advantages compared to the free space setup. The effect of the waveguide cross section on the interaction between the waveguide mode and the sample can be expressed by a factor, called interaction factor. In this article, a new formulation for the interaction factor is derived. It is shown that this factor is inversely proportional to the energy velocity of the waveguide mode.  
\end{abstract}

\maketitle 

\section*{Introduction}
Absorption spectroscopy, which is the study of electromagnetic wave absorption due to its interaction with a material sample, has extensive applications; including identification of organic and inorganic
compounds, polymers, and biological samples \cite{1}. In the conventional absorption spectroscopy setup, electromagnetic wave passes through a material sample in free space. If electromagnetic wave passes through a material sample with power absorption coefficient of $\alpha_{s}$, then based on the Bear-Lambert law \cite{2}:
\begin{equation}
P=P_{0}e^{-\alpha_{s} L}
\end{equation}
in which $P_{0}$ is the input power of the electromagnetic wave, and $P$ is the output power, as shown in Fig. 1 (a). In the free space spectroscopy setup, a large amount of the sample material is required to have a measurable interaction between the sample and the electromagnetic wave. This setup is also very bulky and difficult to implement in a miniaturized integrated package. Waveguide-based absorption spectroscopy setup, in which the signal is passing through a low-loss waveguide instead of free space, and the waveguide is immersed in the sample material [as shown in Fig. 1(b)] is a promising candidate to overcome the difficulties with the free space scheme.
\begin{figure}[h!]
\centering
\includegraphics[width=10cm]{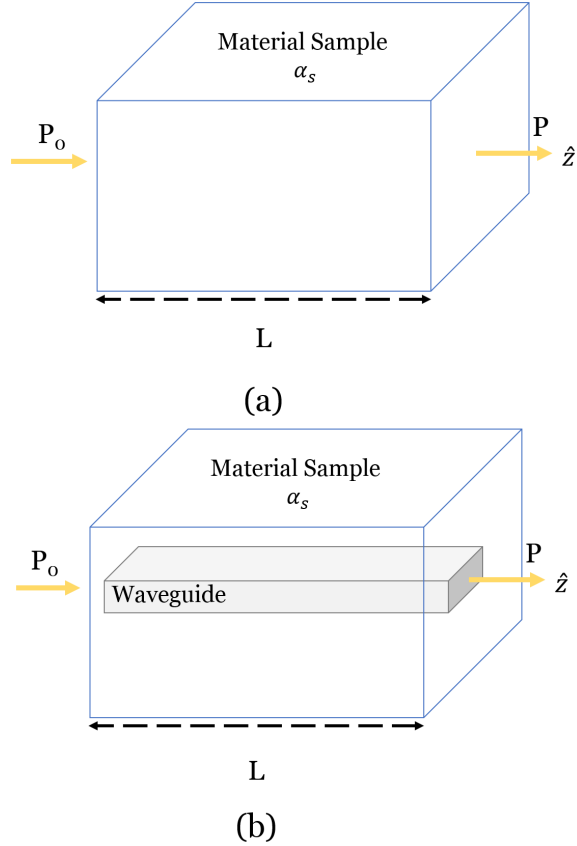}
\caption{Electromagnetic wave passing through a lossy material sample; (a) Free space setup, (b) Waveguide-based setup}
\end{figure}
In the waveguide-based absorption spectroscopy setup, not all of the modal power interacts with the material sample, since a part of the modal field is confined to the waveguide and only the part of the field which is outside of the guiding region passes through the sample. In this case, Beer-Lambert law should be modified to \cite{3}:
\begin{equation}
P=P_{0}e^{-(\alpha_{WG}+\Gamma\alpha_{s}) L}
\end{equation}
in which, $\alpha_{WG}$ is the loss of the waveguide, and $\Gamma$ is the interaction factor denoting the interaction between the waveguide mode and the material sample. In the conventional free space setup, $\Gamma=1$, as all of the beam power transmits through a homogenous sample. In the waveguide-based setup, the interaction factor depends on the cross section of the waveguide and the modal field distribution. In \cite{2} a formulation for the interaction factor is presented when electromagnetic wave passes through a periodic structure. Although same formulation has been used for waveguide-based setup \cite{4,5,6,7}, it is shown in this article that this form is not sufficiently accurate for the waveguide-based cases.\\

In this article, Poynting’s theorem is used to derive the interaction factor in a waveguide-based setup. We will then discuss how the derived expression will be converted to the formulation in \cite{2} in the case of the periodic structure.

\section*{Formulation}
Without loss of generality, we consider a free standing dielectric channel waveguide, shown in Fig. 2. The waveguide is assumed to be immersed in a lossy material sample. Poynting’s theorem is applied to a cylindrical region $V$, enclosed by surface the $S$. Based on Poynting’s theorem, we can write:
\begin{equation}
P_{s}=P_{e}+\overline{P}_{d}+j2\omega(\overline{W}_{m}-\overline{W}_{e})
\end{equation}
\begin{figure}[t!]
\centering
\includegraphics[width=14cm]{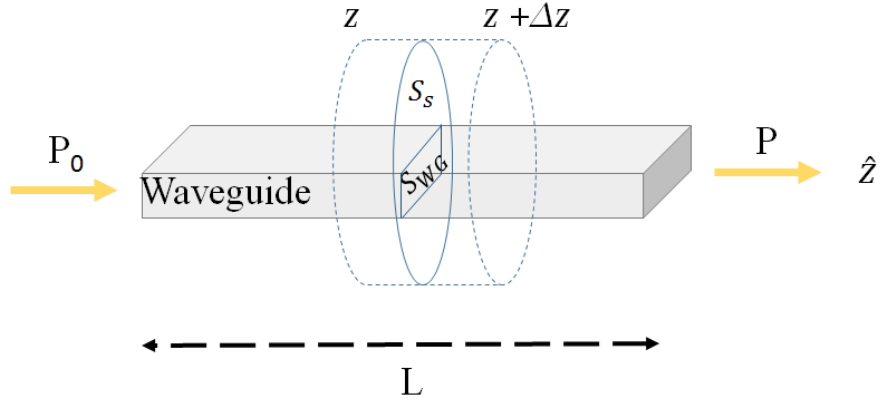}
\caption{Free standing dielectric channel waveguide}
\end{figure}
where, $P_{s}$ is the complex supplied power by the sources inside $V$. $P_{e}$ is the complex power exiting volume $V$:
\begin{equation}
P_{e}=\frac{1}{2}\int_{S}(\textbf{\emph{E}}\times\textbf{\emph{H}}^{*})\cdot\hat{\textbf{\emph{n}}}dS
\end{equation}

$\overline{P}_{d}$ is the time average dissipated power:

\begin{equation}
\overline{P}_{d}=\frac{1}{2}\int_{V}\sigma\left | \textbf{\emph{E}} \right |^{2}dV
\end{equation}

$\overline{W}_{e}$ and $\overline{W}_{m}$ are the time averaged stored energies in the electric and magnetic fields. For the cylinder shown in Fig. 2, as there is no electromagnetic source inside the cylinder, $P_{s}=0$. $\overline{W}_{e}$ and $\overline{W}_{m}$  are equal in a waveguide mode \cite{8}. In this case the Poynting’s theorem will be simplified to:

\begin{equation}
P_{e}+\overline{P}_{d}=0
\end{equation}

In general, integration in (4) should be applied over all faces of the cylinder. Since in an ideal waveguide there is no radiation loss, the integration will be non-zero only on the circular disks.

\begin{equation}
\begin{multlined}
P_{e}=-\frac{1}{2}Re\left\{{\int_{S,z+\Delta z}(\textbf{\emph{E}}\times\textbf{\emph{H}}^{*})\cdot\hat{\textbf{\emph{z}}}dS - \int_{S,z}(\textbf{\emph{E}}\times\textbf{\emph{H}}^{*})\cdot\hat{\textbf{\emph{z}}}dS} \right\}\\
=-P(z+\Delta z)+P(z)=-\Delta z\frac{dP(z)}{dz} \\
=\Delta z\frac{d}{dz}\left\{\frac{1}{2}Re\int_{S}(\textbf{\emph{E}}\times\textbf{\emph{H}}^{*})\cdot\hat{\textbf{\emph{z}}}dS\right\}=-\Delta z\alpha\frac{1}{2}Re\int_{S}(\textbf{\emph{E}}\times\textbf{\emph{H}}^{*})\cdot\hat{\textbf{\emph{z}}}dS
\end{multlined}
\end{equation}

wherein $P(z)$ is the transmitted modal power passing through the cross section $S$ at $z$, and is given by:

\begin{equation}
P(z)=P_{0}e^{-\alpha z}
\end{equation}

wherein $\alpha$ is the mode attenuation constant and $P_{0}$ is the input power. The dissipated power relation (5), can rewritten as:

\begin{equation}
\overline{P}_{d}=\frac{1}{2}\int_{S}\sigma\left | \textbf{\emph{E}} \right |^{2}dS \int_{0}^{\Delta z}e^{-\alpha z} dz
\end{equation}

\begin{equation}
\int_{0}^{\Delta z}e^{-\alpha z} dz=\frac{1-e^{-\alpha\Delta z}}{\alpha}=\Delta z  (for \Delta z\rightarrow 0)
\end{equation}

Also, the integration over the cross section $S$ can be divided to the integration over the cross section of the guiding region of the waveguide,$S_{WG}$, and cross section of the material sample, $S_{s}$. Therefore:

\begin{equation}
\overline{P}_{d}=\Delta z \frac{1}{2}\int_{S}\sigma\left | \textbf{\emph{E}} \right| ^{2}dS=\Delta z \frac{1}{2}(\sigma_{WG}\int_{S_{WG}}\left | \textbf{\emph{E}} \right| ^{2}dS+\sigma_{s}\int_{S_{s}}\left | \textbf{\emph{E}} \right| ^{2}dS)
\end{equation}

On the right hand side of (11), the first term is the loss of the guiding region of the waveguide, and the second term is the loss of the material sample. Using (6), (7), and (11), the mode attenuation constant will be:

\begin{equation}
\alpha=\sigma_{WG}\frac{\int_{S_{WG}}\left | \textbf{\emph{E}} \right| ^{2}dS}{Re\int_{S}(\textbf{\emph{E}}\times\textbf{\emph{H}}^{*})\cdot\hat{\textbf{\emph{z}}}dS}+\sigma_{s}\frac{\int_{S_{s}}\left | \textbf{\emph{E}} \right| ^{2}dS}{Re\int_{S}(\textbf{\emph{E}}\times\textbf{\emph{H}}^{*})\cdot\hat{\textbf{\emph{z}}}dS}
\end{equation}

The power absorption coefficient of the material sample, $\alpha_{s}$, is related to the complex relative permittivity as:

\begin{equation}
\alpha_{s}=2k_{0}Im\sqrt{\varepsilon_{r_{s}}^{'}+j\varepsilon_{r_{s}}^{"}}
\end{equation}

If the material sample is very low-loss, then:

\begin{equation}
\sqrt{\varepsilon_{r_{s}}^{'}+j\varepsilon_{r_{s}}^{"}}\approx \sqrt{\varepsilon_{r_{s}}^{'}}(1+j\frac{\varepsilon_{r_{s}}^{"}}{2\varepsilon_{r_{s}}^{'}})
\end{equation}

Therefore:

\begin{equation}
\sigma_{s}=\omega\varepsilon_{0}\varepsilon_{r_{s}}^{"}=\frac{\alpha_{s}}{\eta_{s}}
\end{equation}

in which $\eta_{s}=\sqrt{\frac{\mu_{s}}{\varepsilon_{s}}}$ is the characteristic impedance of the material sample. Therefore:

\begin{equation}
\sigma_{s}\frac{\int_{S_{s}}\left | \textbf{\emph{E}} \right| ^{2}dS}{Re\int_{S}(\textbf{\emph{E}}\times\textbf{\emph{H}}^{*})\cdot\hat{\textbf{\emph{z}}}dS}=\frac{\alpha_{s}}{\eta_{s}}\frac{\int_{S_{s}}\left | \textbf{\emph{E}} \right| ^{2}dS}{Re\int_{S}(\textbf{\emph{E}}\times\textbf{\emph{H}}^{*})\cdot\hat{\textbf{\emph{z}}}dS}
\end{equation}

Using (2), then interaction factor $\Gamma$ will be:

\begin{equation}
\Gamma=\frac{1}{\eta_{s}}\frac{\int_{S_{s}}\left | \textbf{\emph{E}} \right| ^{2}dS}{Re\int_{S}(\textbf{\emph{E}}\times\textbf{\emph{H}}^{*})\cdot\hat{\textbf{\emph{z}}}dS}
\end{equation}

This relation can be converted to a more comprehensive form as following. The total stored energy in the material sample per unit length along the waveguide ($z$-direction) is:

\begin{equation}
W=\frac{1}{2}{\int_{S_{s}}\varepsilon_{s}\left | \textbf{\emph{E}} \right| ^{2}dS}
\end{equation}

The interaction factor in (7) can then be rewritten as:

\begin{equation}
\Gamma=\frac{1}{\eta_{s}\varepsilon_{s}}\frac{\frac{1}{2}\int_{S_{s}}\varepsilon_{s}\left | \textbf{\emph{E}} \right| ^{2}dS}{\frac{1}{2}\int_{S}\varepsilon\left | \textbf{\emph{E}} \right| ^{2}dS}\frac{\frac{1}{2}\int_{S}\varepsilon\left | \textbf{\emph{E}} \right| ^{2}dS}{\frac{1}{2} Re\int_{S}(\textbf{\emph{E}}\times\textbf{\emph{H}}^{*})\cdot\hat{\textbf{\emph{z}}}dS}
\end{equation}

In (19), the second fraction is the total average stored energy in the material sample divided by the total average stored energy in the both material sample and guiding region of the waveguide, both per unit length of propagation direction ($z$). This term can be defined as a filling factor, representing the percentage of the modal stored energy is in the sample region.

\begin{equation}
f=\frac{\frac{1}{2}\int_{S_{s}}\varepsilon_{s}\left | \textbf{\emph{E}} \right| ^{2}dS}{\frac{1}{2}\int_{S}\varepsilon\left | \textbf{\emph{E}} \right| ^{2}dS}
\end{equation}

This expression can also be expressed in terms of the waveguide energy velocity, which is defined as \cite{8}:

\begin{equation}
v_{en}=\frac{P_{z}}{W}=\frac{\frac{1}{2} Re\int_{S}(\textbf{\emph{E}}\times\textbf{\emph{H}}^{*})\cdot\hat{\textbf{\emph{z}}}dS}{\frac{1}{2}\int_{S}\varepsilon\left | \textbf{\emph{E}} \right| ^{2}dS}
\end{equation}

Therefor (19) can be rewritten as:

\begin{equation}
\Gamma=\frac{1}{\eta_{s}\varepsilon_{s}}f\frac{1}{v_{en}}
\end{equation}

Furthermore:

\begin{equation}
\frac{1}{\eta_{s}\varepsilon_{s}}=\frac{1}{\sqrt{\frac{\mu_{s}}{\varepsilon_{s}}}\varepsilon_{s}}=\frac{1}{\sqrt{\mu_{s}\varepsilon_{s}}}=v_{s}
\end{equation}

where $v_{s}$ is the speed of light in the sample material medium. Therefore (22) can be written in its most concise form as:

\begin{equation}
\Gamma=f\frac{v_{s}}{v_{en}}
\end{equation}

In 
\cite{2} a similar formula has been derived, wherein the energy velocity, $v_{en}$, is replaced by group velocity, $v_{g}$, for a periodic structure. Although in some special cases, including a periodic structure, $v_{g}=v_{en}$ and therefore (24) and the expression derived in \cite{2} become identical, the rigorous expression for the interaction factor for a general waveguide is the one given in (24).

\section*{Conclusion}

Interaction factor in the waveguide-based absorption spectroscopy defines the impact of the waveguide structure and the modal field distribution on the interaction between the waveguide mode and the lossy surrounding material sample. Using Poynting’s theorem, we proved that this factor depends on the speed of the light in the material sample region, a filling factor $f$, and the energy velocity of the
waveguide mode. Although the group velocity of the waveguide has been used in literature to calculate the interaction factor, based on the rigorous formulation presented here, in general the energy velocity of the waveguide, $v_{en}$, should be used in the expression for the interaction factor.


\end{document}